\documentstyle[prl,aps,multicol,epsfig]{revtex}

\begin{document}
\large

\title{Quantum Information processing by NMR: Preparation of pseudopure states and 
implementation of unitary operations in a single-qutrit system.}
\author{Ranabir Das$^\dagger$, Avik Mitra$^\dagger$, Vijay Kumar S$^{\top}$ and Anil Kumar $^{\dagger \ddagger}$\\
        $^{\dagger}$ {\small \it Department of Physics, Indian Institute of Science, Bangalore, India}\\
        $^{\top}$ {\small \it Department of Electrical Engineering, Indian Institute of Technology, Madras, India}\\
        $^{\ddagger}$ {\small \it Sophisticated Instruments Facility, Indian Institute of Science, Bangalore, India}\\}

\maketitle
\vspace{0.5cm}
\begin{abstract}
Theoretical Quantum Information Processing (QIP) has matured from the use of qubits to the 
use of qudits (systems having states $>$ 2). Where as most of the experimental implementations 
have been performed using qubits, little experimental work has been carried out using qudits as yet. In this paper we demonstrate
experimental realization of a qutrit system by nuclear magnetic resonance (NMR), utilizing deuterium (spin-1)
nuclei partially oriented in liquid crystalline phase.
Preparation of pseudopure states and implementation of unitary operations are demonstrated in
this single-qutrit system, using transition selective pulses.\\\\
\end{abstract}

\section{Introduction}
 A future quantum computer has the advantage of simulating physical systems and 
solving certain problems  more efficiently than classical computers \cite{rf,ss,deu,pw,gr}. Classical 
computer use bits as basic units of information whereas quantum computers 
use qubits (two-level systems) as basic units of information. Unlike bits, qubits can exist in 
coherent superposition of the two basic states $\vert 0\rangle$ and $\vert 1\rangle$. 
Moreover quantum-mechanical systems can be 'entangled'. Entanglement is purely 
a quantum-mechanical property which has no classical analogue. Two quantum systems when entangled 
show non-local correlations which cannot be defined by classical statistical predictions. 
 However, such quantum mechanical behavior is restricted not only to qubits but exists 
in all higher dimensional systems called qudits \cite{qud,quent}. 
The degree of non-local 
correlations are in fact greater in case of entangled qutrits (three-level systems)
 than entangled qubits \cite{comm}. Several cryptographic protocols have been devised using 
qutrits which are argued to be highly secure against symmetric attacks\cite{cryp2,cryp,cryp1}. 
Recently it has also been demonstrated that qutrits can be useful for certain purposes of quantum simulation \cite{simu}, 
quantum computations \cite{monty,qgate,qst} and quantum communication 
\cite{byzan,ent,bowen,dense}. Among the various physical systems
on which quantum simulations and algorithms have been experimentally realized,
 nuclear magnetic resonance (NMR) has proved to be the most successful to date \cite{dg,na,ernst,jo,free,kd,rd}.
 In most of the studies coupled spin-1/2 nuclei have been used which simulate multiqubit systems.
 Quadrupolar nuclei have also been used in NMR to mimic qubit systems, whereby spin-3/2 and 7/2 nuclei 
act as 2  and  3-qubit systems respectively \cite{fun,mulf,sim,ne,mur,rd1}. In 
this work we realize a single qutrit system using nucleus of a deuteron (spin-1) oriented 
in a liquid crystal matrix. The quadrupole moment of spin-1 coupled to the electric field gradient 
in an anisotropic environment lifts the degeneracy of the various dipolar transitions in a strong 
magnetic field, allowing the three level system to be treated as a "qutrit". 
We prepare pseudopure states and perform all the
 possible one-to-one reversible operations on this qutrit system.  To the best of our knowledge this is 
the first implementation of quantum information processing in qutrit system by NMR spectroscopy. 
It is worth mentioning that qutrit systems have also been realized in trapped-ions recently \cite{ion}.
            
    A qutrit has three states $\vert 0 \rangle$, $\vert 1 \rangle$ and  $\vert 2 \rangle$. A natural choice 
for qutrit is a nucleus with three energy levels. Such a nucleus commonly used in 
NMR spectroscopy is deuterium ($^2$H) which has nuclear spin-1 and significant quadrupole moment. 
The energy levels of deuterium in a strong magnetic field are schematically given in Fig. 1(a) along with the 
equilibrium populations. In a strong magnetic field (B$_o$) and an isotropic environment the quadrupolar coupling 
is averaged to zero; consequently, the energy levels of deuterium are equispaced. Hence the two single quantum transitions 
( $\vert 0 \rangle \leftrightarrow \vert 1 \rangle$ and $\vert 1 \rangle \leftrightarrow \vert 2 \rangle$)    
overlap leading to indistinguishibility and causing problems with addressability. However, when introduced in a liquid 
crystalline matrix, the system becomes anisotropically oriented with respect to B$_o$, 
yielding finite quadrupolar coupling. The resonance frequencies of the
 two single quantum transitions then become non-degenerate, giving well resolved spectra (Fig.1(a)). 
 The Hamiltonian of a quadrupolar nucleus partially oriented in a liquid crystalline matrix, in the presence
of a large magnetic field $B_0$ and  having a first-order quadrupolar coupling, is given by \cite{khel};
\begin{eqnarray}
\mathcal{H}=\mathcal{H} _Z+\mathcal{H}_Q &=&-\omega_0 I_z + \frac{e^2qQ}{4I(2I-1)} (3I^2_z-I^2)S \nonumber \\
   &=& - \omega_0 I_z+\Lambda (3I^2_z-I^2),
\end {eqnarray}
 where $\omega_0=\gamma B_0$ is the resonance frequency, $\gamma$ being the gyro-magnetic ratio,
 $S$ is the order parameter at the site of the nucleus,  $e^2qQ$ is the quadrupolar coupling and
$\Lambda = e^2qQS/(4I(2I-1))$ is its effective value.
Though $e^2qQ$ is of the order of several MHz,
 a small value for the order parameter ($S$) converts the effective quadrupolar coupling `$\Lambda$' into several Hz.
 The sample used in this study is a lyotropic liquid crystal comprising of 25.6$\%$ potassium laurate, 
6.24$\%$ decanol and 68.16$\%$ deuterated water 
(D$_2$O) \cite{liq}. The system exhibits oriented nematic laminar phase at 300K \cite{liq}. The deuterium signal 
splits into a doublet of 240 Hz separation due to partially averaged quadrupolar interaction (Fig.1(a)). 
All experiments were performed at 
a magnetic field of 11.7 Telsa at which the resonance frequency of deuterium is 76.75 MHz.
 We have utilized this qutrit system for creation of pseudopure states and 
implementation of several one-to-one reversible operations.    
                            
\section{Pseudopure state}
   In a pure state each member of an ensemble can be described by the same vector \cite{dg,na}.
 In equilibrium at a finite temperature 
and magnetic field, the ensemble of nuclear spins will be in a mixed state whose density matrix is of the form 
\begin{eqnarray}
\sigma=c\bf{I}+ \sigma_{\Delta}
\end{eqnarray}   
 where $\bf{I}$ is an identity matrix, c is a constant whose value is approximately $10^6$ for 
nuclear spins at room temperature and $\sigma_{\Delta}$ is the traceless deviation density matrix. 
Though the identity matrix does not evolve under radio frequency pulses and system Hamiltonian, the 
deviation part does evolve and emits the entire signal. In equilibrium both $\sigma$ and $\sigma_{\Delta}$ are 
in mixed state. While it is requires extreme conditions of a very low temperature to create a pure $\sigma$, 
the deviation density matrix can be easily transformed, such that it behaves like a pure state. By using r.f. pulses 
and magnetic field gradients, the populations of all the levels except one can be equalized, in which case it behaves 
like a pure state. The state of the system when its deviation matrix corresponds to a pure state, is known as a 
pseudopure state \cite{dg,na}. Pseudopure states (pps) have been proposed and implemented in qubit systems 
using several experimental schemes by NMR \cite{dg,na,ernst,jo,fun,mulf,ne,rd1}. Here we create  
the various pseudopure states in a single-qutrit system, by spatial averaging technique using transition 
selective pulses \cite{ernst,ne}.

 Transition selective pulses are long duration, low power r.f pulses applied at the 
resonant frequency between two energy levels connected by $\Delta$m=$\pm$1, where m is the spin magnetic quantum number. 
Such a pulse  excites the single quantum transition between those levels and has been used earlier 
 for quantum information processing in qubit systems\cite{free,kd,ne,mur,rd1}. We have used these selective 
pulses for quantum information processing in qutrit systems.
A selective $\pi$ pulse tuned at the resonant frequency between two energy levels interchanges
 the populations between them \cite{rd1}. Similarly a selective $\pi/2$ pulse between two energy levels equilibrate the 
populations between the two levels and creates coherences which can be subsequently destroyed by a gradient pulse \cite{rd1}. 
Starting from thermal equilibrium populations of Fig 1(a), a $\pi/2$ pulse applied on the transition 
$\vert 1 \rangle \leftrightarrow \vert 2\rangle$ ($(\pi/2)^{\vert 1 \rangle \leftrightarrow \vert 2\rangle}$) followed by a 
gradient to equalize the populations of $\vert 1 \rangle$ and $\vert 2 \rangle$,  created $\vert 0 \rangle$ 
pseudopure state (Fig 1(b)). $\vert 1 \rangle$ pps is created by a $(\pi/2)^{\vert 1 \rangle \leftrightarrow \vert 2\rangle}$
 followed by a gradient pulse to equalize the populations of $\vert 1 \rangle$ and $\vert 2 \rangle$ and then a 
$(\pi)^{\vert 1 \rangle \leftrightarrow \vert 0\rangle}$ pulse to interchange the populations between $\vert 1 \rangle$ and 
$\vert 0 \rangle$ (Fig 1(c)). Finally another gradient with different strength of the earlier was applied to destroy all coherences 
created by the imperfection of pulses. The $-\vert 2 \rangle$ pps was prepared by a 
$(\pi/2)^{\vert 1 \rangle \leftrightarrow \vert 0\rangle}$ pulse followed by  
a gradient. The created pseudopure states were detected using a final non-selective small-angle (5$^o$) pulse for all the experiments. 
This small-angle pulse, in linear approximation,  
converts the population differences into single quantum transitions whose intensities are proportional to the population 
differences of the two involved levels only. The experimental spectra given in Fig. 1 confirm the preparation of the various pseudopure 
states, in this qutrit system.    

\section{Single-qutrit operations}
 There are six possible one-to-one reversible operations in a single qutrit system. These operations corresponds to six U(3) operations.
 Starting from equilibrium we implement all these six unitary operations.  The input, the output and the operators 
corresponding to the different operations are listed in Table 1 along with the pulses required to implement them. 
 The pulses described in third column of table 1 are to be applied from left to right. Each one of the 
 $(\pi)$ pulses correspond to a SU(2) operation, since it operates on  any two of the three states. 
 U1 corresponds to identity operation and requires no pulses. U2 exchanges the states of $\vert 1 \rangle$ 
and $\vert 2 \rangle$, and is implemented by a $(\pi)^{\vert 1 \rangle \leftrightarrow \vert 2\rangle}$ pulse.
As predicted earlier, the U(3) operations in a single qutrit can be split into 
series of SU(2) operations \cite{qgate,ion,wh1}. Hence we note that while U2 and U3 are implemented with one SU(2) operation, 
 U5 and U6 required a series of two SU(2) operations.  U5 and U6 were implemented by 
$(\pi)^{\vert 1 \rangle \leftrightarrow \vert 2\rangle} (\pi)^{\vert 0 \rangle \leftrightarrow \vert 1\rangle}$ 
and $(\pi)^{\vert 0 \rangle \leftrightarrow \vert 1\rangle} (\pi)^{\vert 1 \rangle \leftrightarrow \vert 2\rangle}$ 
respectively.
  While it is evident that the U4 operation can be performed by three SU(2) operations in the series 
$(\pi)^{\vert 1 \rangle \leftrightarrow \vert 2\rangle}
(\pi)^{\vert 0 \rangle \leftrightarrow \vert 1\rangle}
(\pi)^{\vert 1 \rangle \leftrightarrow \vert 2\rangle}$ or 
$(\pi)^{\vert 0 \rangle \leftrightarrow \vert 1\rangle}
(\pi)^{\vert 1 \rangle \leftrightarrow \vert 2\rangle}
(\pi)^{\vert 0 \rangle \leftrightarrow \vert 1\rangle}$, 
the same can also be performed by a non-selective $\pi$-pulse. We have implemented the later for the ease of the experiment.

 These single-qutrit unitary operations will be useful for quantum information in qutrits. Specially the last three
operations of U4-U6 are applied in quantum computation \cite{qgate,ion,wh1}, quantum games \cite{monty} and  quantum communication
\cite{bowen,dense}. U4 is used as NOT operation in unary algebra \cite{net}, while U5 and U6 are used as
rotate-up and rotate-down operations in unary algebra \cite{net}.
Both for preparation of pseudopure states and implementation of unitary operations, 
Gaussian shaped pulses with a duration of 6ms were used as selective pulses.  
After each selective pulse a sine-bell shaped gradient pulse was applied to destroy any coherences created by imperfection of pulses.
Since we start from thermal equilibrium, the result after each unitary operation is encoded in the output populations of the different 
states. The output populations were measured by using a non-selective flip-angle (5$^o$) pulse.   
The experimental spectrum after implementation of each operation is given in Fig.2. The observed intensities 
are within 5$\%$ of the expected intensities.
\section{Conclusion}
 Here we have demonstrated creation of pseudopure states and implementation of unitary operations in a single qutrit system 
using a deuterium nuclei oriented in liquid crystalline matrix. This implementation extends the range of systems 
useful for quantum information processing by NMR. These systems can now be coupled with other nuclei to yield larger 
systems for quantum information processing. Efforts are ongoing to create such systems and implement quantum 
computation and information processing in them.

 The authors thank H.S. Vinay Deepak and K.V. Ramanathan
for useful discussions. The use of DRX-500 NMR spectrometer funded by the Department of
Science and Technology, New Delhi, at the Sophisticated
Instruments Facility, Indian Institute of Science, Bangalore, is also gratefully acknowledged.
  

\pagebreak

TABLE 1. Unitary operations and their pulse schemes in a single-qutrit system.
\\\\
\hspace*{4cm}
\begin{tabular}{|c|c|c|c|c|}\hline
\quad Operation \quad &  \quad In \quad & \quad Out  & \quad Operator \quad & Pulse \cr \hline

&&&&\cr
U1 & $\matrix{ \vert 0\rangle \cr \vert 1\rangle \cr \vert 2\rangle }$ &
$\matrix{ \vert 0\rangle \cr \vert 1\rangle \cr \vert 1\rangle }$ &$\pmatrix{1~~ 0~~ 0\cr 0~~ 1~~ 0\cr 0~~ 0~~ 1}$ & no pulse \cr
&&&&\cr \hline
&&&&\cr
U2 & $\matrix{\vert 0\rangle \cr \vert 1\rangle \cr \vert 2\rangle }$ &
$\matrix{\vert 0\rangle \cr \vert 2\rangle \cr \vert 1\rangle }$ &$\pmatrix{1~~ 0~~ 0\cr 0~~ 0~~ 1\cr 0~~ 1~~ 0}$
& $(\pi)^{\vert 1 \rangle \leftrightarrow \vert 2\rangle}$ \cr &&&&\cr \hline
&&&&\cr
U3 & $\matrix{\vert 0\rangle \cr \vert 1\rangle \cr \vert 2\rangle }$ &
$\matrix{\vert 1\rangle \cr \vert 0\rangle \cr \vert 2\rangle}$ &$\pmatrix{0~~ 1~~ 0\cr 1~~ 0~~ 0\cr 0~~ 0~~ 1}$ &
$(\pi)^{\vert 0 \rangle \leftrightarrow \vert 1\rangle}$ \cr &&&&\cr \hline
&&&&\cr
U4 & $\matrix{\vert 0\rangle \cr \vert 1\rangle \cr \vert 2\rangle}$ &
$\matrix{\vert 2\rangle \cr \vert 0\rangle \cr \vert 0\rangle }$ &$\pmatrix{0~~ 0~~ 1\cr 0~~ 1~~ 0\cr 1~~ 0~~ 0}$
& non-selective $(\pi)$ \cr &&&&\cr \hline
&&&&\cr
U5 & $\matrix{\vert 0\rangle \cr \vert 1\rangle \cr \vert 2\rangle}$ &
$\matrix{\vert 1\rangle \cr \vert 2\rangle \cr \vert 0\rangle}$ &$\pmatrix{0~~ 1~~ 0\cr 0~~ 0~~ 1\cr 1~~ 0~~ 0}$ &
$(\pi)^{\vert 1 \rangle \leftrightarrow \vert 2\rangle} (\pi)^{\vert 0 \rangle \leftrightarrow \vert 1\rangle}$
\cr &&&&\cr \hline
&&&&\cr
U6 & $\matrix{\vert 0\rangle \cr \vert 1\rangle \cr \vert 2\rangle }$ &
$\matrix{\vert 2\rangle \cr \vert 0\rangle \cr \vert 1\rangle}$ &$\pmatrix{0~~ 0~~ 1\cr 1~~ 0~~ 0\cr 0~~ 1~~ 0}$ &
$(\pi)^{\vert 0 \rangle \leftrightarrow \vert 1\rangle} (\pi)^{\vert 1 \rangle \leftrightarrow \vert 2\rangle}$ \cr &&&&\cr\hline
\end{tabular}
\pagebreak
\hspace{6cm}\large{FIGURE CAPTIONS}
\\\\
FIG. 1: Preparation of pseudopure states in a qutrit system. 
(a) Energy level diagram of a spin-1 nucleus oriented in a liquid crystal matrix. The different
spin states can be labeled as qutrit states. The equilibrium deviation populations of different states 
under high-field high-temperature approximation are schematically shown by filled circles on the right hand side.
 The equilibrium spectrum of deuterium ($^2$H) contains a 
doublet of equal height and separated by an effective quadrupolar coupling ($\Lambda$) of  240 Hz. 
The $\vert 0\rangle$, $\vert 1\rangle$ and $\vert 2\rangle$ pseudopure states are given respectively in 
Figures  1(b), 1(c) and 1(d). The various pseudopure states were obtained starting from equilibrium populations 
by using the pulse sequence given in each figure, followed by a crusher gradient pulse. All the spectra were obtained 
by using a small angle ($5^o$) non-selective pulse.  
The populations after creation of pseudopure states in each case are also schematically shown. 
The length of the transition selective pulses was kept fixed as 6 ms, and
 the strength of applied r.f. field was calibrated for selective ($\pi$) and ($\pi/2$) pulses respectively as 83.3 Hz and 41.67 Hz.  
The expected and observed (in brackets) intensity ratios of the deuterium doublet  are (a) 1:1 [0.99:1], (b) 0:1.5  [0.01:1.5], 
(c) 1.5:-1.5 [1.5:-1.42] and (d) 1.5:0 [1.5:-0.01].    
\\\\

FIG. 2: Single qutrit operations in deuterium oriented in liquid crystal matrix. All the U(3) operations are 
implemented by the pulse sequence given in the last column of Table 1, 
using  selective and non-selective $\pi$ pulses of length 6 ms and 42 $\mu$s respectively.
 The final populations were measured by small angle ($5^o$) non-selective pulse. The  expected and
observed (in brackets) intensity ratios of the deuterium doublet are (a) 1:1 [0.99:1], (b) 1:1 [0.99:1], (c) -1:2 [-0.96:2],
(d) 2:-1 [2:-1.02], (e) -1:-1 [-0.98:-1], (f) 1:-2 [1.04:-2] and  (g) -2:1 [-2:1.06]. 

\pagebreak
\begin{figure}
\epsfig{file=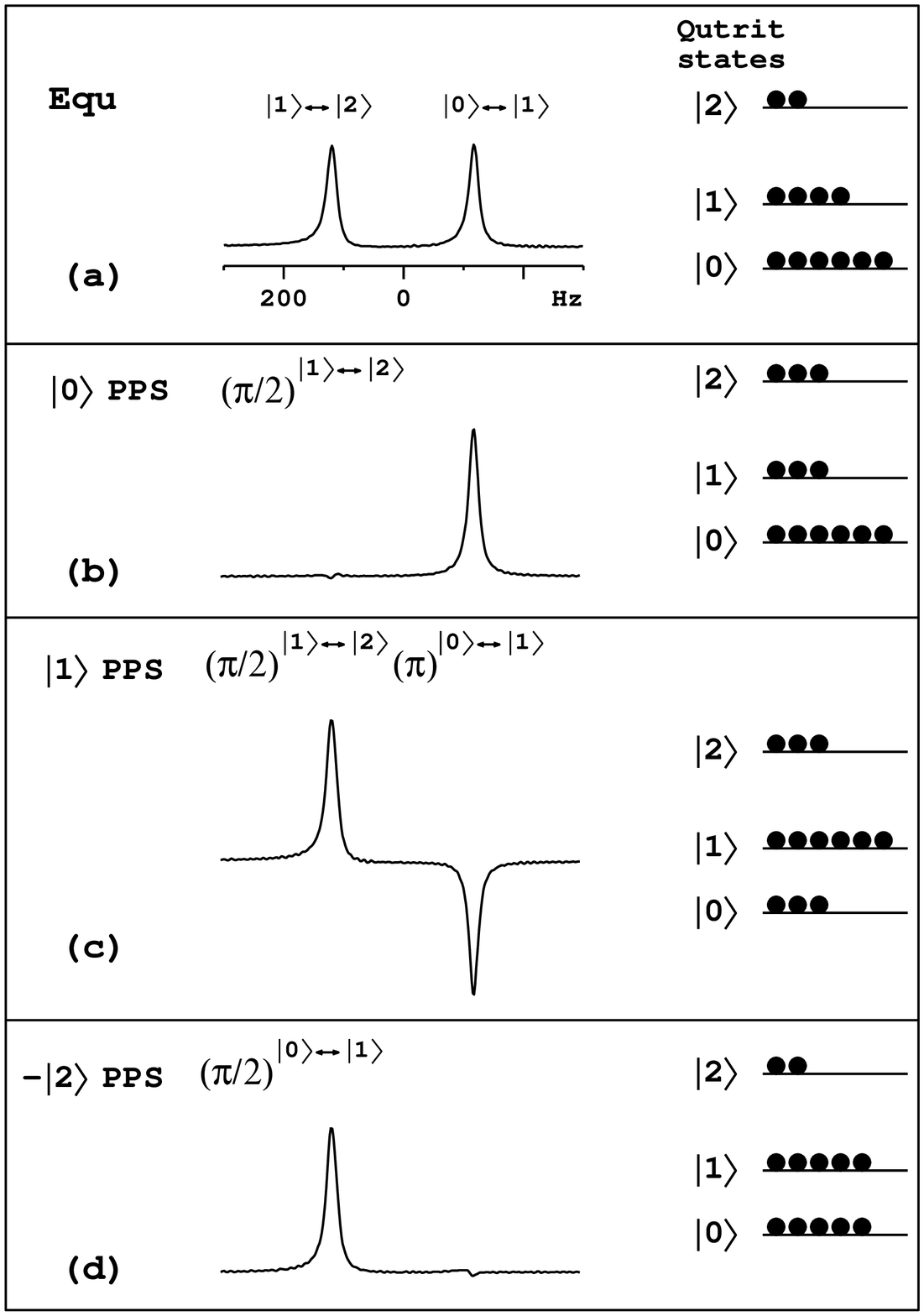,height=20cm}
\end{figure}
\hspace{6cm}
{\Large Figure 1}
\pagebreak
\begin{figure}
\epsfig{file=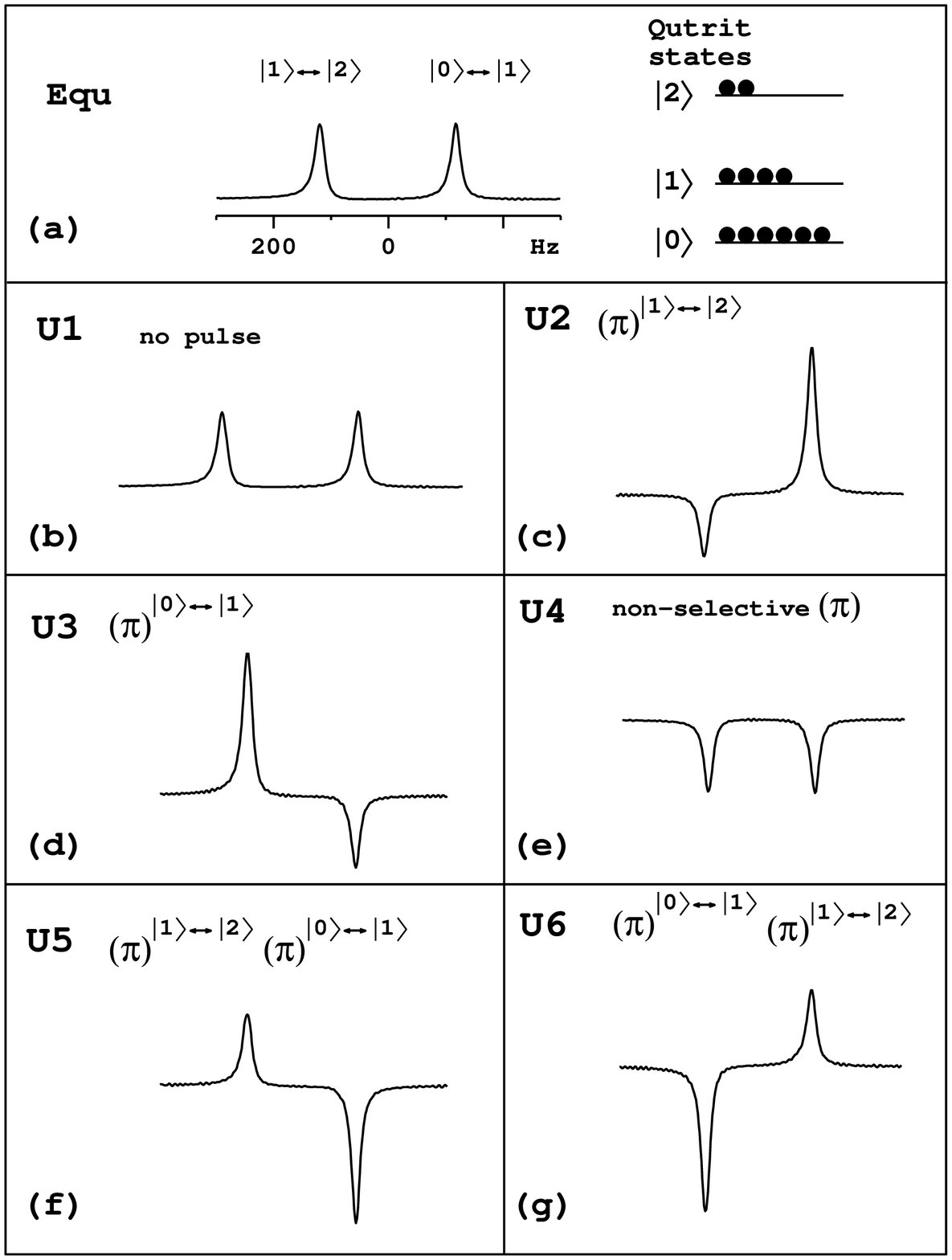,height=22cm}
\end{figure}
\hspace{6cm}
{\Large Figure 2}


\begin{thebibliography}{99}
\bibitem{rf} R.P. Feynman, Int J. Theor. Phys. 21, (1982) 467.
\bibitem{ss} S. Lloyd, Science. 273, (1996) 1073.
\bibitem{deu} D. Deutsch and R. Jozsa, Proc. R. Soc. London A 439, (1992) 553.
\bibitem{pw} P.W.Shor,
{\em Proceedings of the 35th Annual Symposium in Foundations of Computer Science,
Santa Fe, NM, 1994 }(IEEE Computer Society Press, Los Alamitos, CA,1994).
\bibitem{gr} L.K. Grover, Phys. Rev. Lett. 79, (1997) 325.
\bibitem{qud} Carlton M. Caves and Gerard J. Milburn, Opt. Commmun. 179, 439 (2000).
\bibitem{quent} Vivien. M. Kendon, Karol Zyczkwoski and W. J. Munro Phys. Rev. A 66, 062310 (2002). 
\bibitem{comm} D. Kaszlikowski, D. Gosal, E. J. Ling, L. C. Kwek, M. Zukowski, and C. H. Oh, Phys. Rev. A 66, 032103 (2002).
\bibitem{cryp2} D. Bruss and C. Macchiavello, Phys. Rev. Lett. 88, 127901 (2002);
N. J. Cerf,1,2 M. Bourennane, A. Karlsson, and N. Gisin, Phys. Rev. Lett. 88, 127902 (2002).
\bibitem{cryp} T. Durt, N. J. Cerf, N. Gisin, and M. Zukowski, Phys. Rev. A 67, 012311 (2003).
\bibitem{cryp1} D. Kaszlikowski, D. K. L. Oi, M. Christandl, K. Chang, A. Ekert, L. C. Kwek,
\bibitem{simu} B. M. Terhal, I. L. Chuang, D. P. DiVincenzo, M. Grassl, and J. A. Smolin, Phys. Rev. A 60, 881-885 (1999).
\bibitem{monty} A. P. Flitney and D. Abbott, Phys. Rev. A 65, 062318 (2002).
\bibitem{qgate} J. Kempe and K. B. Whaley, Phys. Rev. A 65, 052330 (2002).
\bibitem{qst} R. T. Thew, K. Nemoto, A. G. White, and W. J. Munro, Phys. Rev. A 66, 012303 (2002).
\bibitem{byzan} M. Fitzi, N. Gisin, and U. Maurer, Phys. Rev. Lett. 87, 217901 (2001).
\bibitem{ent} M. Cinchetti and J. Twamley, Phys. Rev. A 63, 052310 (2001).
\bibitem{bowen} G. Bowen, Phys. Rev. A 63, 022302 (2001).
\bibitem{dense} X. S. Liu, G. L. Long, D. M. Long and F. Li, Phys. Rev. A 65, 022304 (2002).
\bibitem{dg} D.G. Cory, A.F. Fahmy, and T.F. Havel, Proc Natl Acad Sci. USA 94, (1997)  1634.
\bibitem{na} N.A. Gershenfeld and I.L. Chuang, Science. 275, (1997) 350.
\bibitem{ernst} Z.L. Madi, R. Bruschweiler and R.R. Ernst,
One- and two-dimensional ensemble quantum computing in spin Louivelle space,
 J. Chem. Phys. 109, 10603.
\bibitem{jo} J.A. Jones, Prog. Nucl. Mag. Res. Spec. 38, (2001) 325.
\bibitem{free} N. Lindan, H. Barjat, and R. Freeman {\it Chem. Phys. Lett.} {\bf 296}, 61 (1998).
\bibitem{kd} Kavita Dorai, Arvind, Anil Kumar, {\it Phys Rev A.} {\bf 61}, (2000) 042306.
\bibitem{rd} Ranabir Das, T.S. Mahesh, and Anil Kumar, Phys. Rev. A. 67, 062304 (2003).
\bibitem{fun} A.K. Khitrin and B.M. Fung, {\it J. Chem. Phys.} {\bf 112}, 6963 (2000).
\bibitem{mulf} A. Khitrin, H. Sun, and B.M. Fung, {\it Phys. Rev. A} {\bf 63}, 020301(R) (2001).
\bibitem{sim} A. K. Khitrin and B.M. Fung,  {\it Phys. Rev. A} {\bf 64}, 032306 (2001).
\bibitem{ne} Neeraj Sinha, T. S. Mahesh, K.V. Ramanathan, and Anil Kumar,
              J. Chem. Phys. 114, (2001) 4415.
\bibitem{mur}  K.V.R.M. Murali, Neeraj Sinha, T.S. Mahesh, Malcom Levitt, K.V. Ramanathan,
 and Anil Kumar, {\it Phys. Rev. A} {\bf 66}, 022313 (2002).
\bibitem{rd1} Ranabir Das and Anil Kumar, Phys. Rev. A. (accepted).
\bibitem{ion}A. B. Klimov, R. Guzmán, J. C. Retamal, and C. Saavedra,Phys. Rev. A 67, 062313 (2003).
\bibitem{khel} P. Dhiel and C.L. Khetrapal, "NMR-Basic Principles and Progress" Vol.1, Springer-Verlag, New York, 1969.
\bibitem{liq} L. J. Yu and A. Saupe, Phys. Rev. Lett. 45, 1000 (1980).
\bibitem{wh1} J. Kempe, D. Bacon, D. Lidar and  K. B. Whaley, Phys. Rev. A 63, 042307 (2001).
\bibitem{net} http://www.trinary.cc/Tutorial/Algebra/Unary.htm
\end{thebibliography}
\end{document}